%% file: IF2009.tex
\documentclass[letterpaper]{proc}

 \usepackage[dvips]{color}
 \usepackage{amsmath}
 \usepackage[center]{subfigure}
 \usepackage{bm}

 \usepackage{algorithm}
 \usepackage{algorithmic}
 \usepackage{booktabs}

\usepackage[dvips]{graphicx}
\usepackage[english]{babel}
\usepackage[latin1]{inputenc}

\usepackage{amssymb}
\usepackage{url}

\usepackage{cite} 
\usepackage{stfloats}  

\usepackage{times}

\makeatletter
\def\@maketitle{%
  \vbox to 2.3in{%
    \hsize\textwidth
    \linewidth\hsize
    \vspace*{1.5cm}
    \centering
    {\bfseries\LARGE \@title \par}
    \vskip 2em
    {\large \begin{tabular}[t]{c}\@author \end{tabular}\par}
    \vfill}    \vspace*{1.0cm}
}
\renewcommand\section{\@startsection {section}{1}{\z@}%
     {.7\baselineskip plus\baselineskip}{.5\baselineskip}
                                   {\normalfont\Large\bfseries}}
\renewcommand\section{\@startsection {section}{1}{\z@}%
      {.5\baselineskip\@plus.7\baselineskip}{.3\baselineskip}%
                                   {\normalfont\Large\bfseries}}
\renewcommand\subsection{\@startsection{subsection}{2}{\z@}%
       {.5\baselineskip\@plus.7\baselineskip}{.3\baselineskip}%
                                   {\normalfont\large\bfseries}}
\renewcommand\subsubsection{\@startsection{subsubsection}{3}{\z@}%
      {.5\baselineskip\@plus.7\baselineskip}{.3\baselineskip}%
                                     {\normalfont\normalsize\bfseries}}
\makeatother

\renewenvironment{abstract}%
  {\normalfont
    \list{}{\labelwidth0pt
      \leftmargin0pt \rightmargin\leftmargin
      \listparindent\parindent \itemindent0pt
      \parsep0pt
      
    }%
    \item[\hskip\labelsep\bfseries\abstractname\enspace --] \itshape%
}{%
  \endlist}

\newcommand{\keywordsname}{Keywords}
\newenvironment{keywords}%
  {\normalfont
    \list{}{\labelwidth0pt
      \leftmargin0pt \rightmargin\leftmargin
      \listparindent\parindent \itemindent0pt
      \parsep0pt
      }%
    \item[\hskip\labelsep\bfseries\keywordsname:]}{\endlist}
  
\begin{document}

\title{Decision Based Uncertainty Propagation Using Adaptive Gaussian Mixtures}
\author{\begin{tabular}{ccccccc}
\bf Gabriel Terejanu$^{a}$ && \bf Puneet Singla$^{b}$ & &\bf Tarunraj Singh$^b$ & & \bf Peter D. Scott$^a$\\
terejanu@buffalo.edu &&psingla@buffalo.edu && tsingh@buffalo.edu && peter@buffalo.edu
\end{tabular}\\
$^a$\textsl{Department of Computer Science \& Engineering}\\
$^b$\textsl{Department of Mechanical \& Aerospace Engineering}\\
\textsl{University at Buffalo, Buffalo, NY-14260.}} \maketitle

\selectlanguage{english}

\begin{abstract}
Given a decision process based on the approximate probability density function returned by a data assimilation algorithm, an interaction level between the decision making level and the data assimilation level is designed to incorporate the information held by the decision maker into the data assimilation process. Here the information held by the decision maker is a loss function at a decision time which maps the state space onto real numbers which represent the threat associated with different possible outcomes or states. The new probability density function obtained will address the region of interest, the area in the state space with the highest threat, and will provide overall a better approximation to the true conditional probability density function within it. The approximation used for the probability density function is a Gaussian mixture and a numerical example is presented to illustrate the concept.
\end{abstract}

\begin{keywords}
Adaptive Gaussian Sum, Decision Making, Uncertainty Propagation, Expected Loss.
\end{keywords}

%
\section{Introduction}
\label{sec:Introduction}
\input{introduction.tex}

\section{Problem Statement}
\label{sec:Problem}
\input{problem.tex}
\section{Approximation of the Conditional Probability Density Function}
\label{sec:Approximation}
\input{approximation.tex}

\input{loss.tex}
\section{Decision Maker - Data Assimilation Interaction Level}
\label{sec:Method}
\input{method.tex}
\input{meas.tex}
\section{Numerical Results}
\label{sec:Results}
\input{results.tex}

\section{Conclusion}
\label{sec:Conclusion}
\input{conclusion.tex}
\\

\textbf{Acknowledgment:} \textit{This work was supported under Contract No. HM1582-08-1-0012 from ONR.}

\bibliographystyle{plain}
\bibliography{IF2009}

\end{document}

%% file: introduction.tex
Chemical, Biological, Radiological, and Nuclear (CBRN) incidents are rare events but very consequential, which mandates extensive research and operational efforts in mitigating their outcomes. For such critical applications the accuracy in predicting the future evolution of toxic plumes in a timely fashion represents an important part in the Decision Maker (DM) toolbox. Based on these forecasts, decisions can be made on evacuating cities, sheltering or medical gear deployment. Such decisions are taken based on a loss function or region of interest such as the population density in a given area. 

Many research projects try to model the atmospheric transport and diffusion of toxic plumes. While inherently stochastic and highly nonlinear, these mathematical models are able to capture just a part of the dynamics of the real phenomenon and the forward integration yields an uncertain prediction. The decision maker takes actions based on expected loss computed using both the predicted uncertainty and the loss function, which here maps a region of interest in the state space into a threat level, such as the population density in a city. Thus the ability to propagate the uncertainty and errors throughout the dynamic system is of great importance, and the evolution of the probability density function (pdf) has received much attention recently. 

Due to the uncertainty accumulation in integrating the model, the forecasts of the system become less and less useful for the decision maker. Data Assimilation (DA) offers a way to reduce the uncertainty by combining measurements provided by sensors with model prediction in a Bayesian way \cite{Terejanu2007}. This gives an improved situation assessment for the hindcast and nowcast. Unfortunately  the forecast, used to evaluate the impact assessment, is still affected by the accuracy of probability density function evolution. Even in the hindcast and nowcast cases if the sensors provide ambiguous measurements such a quadratic measurement model, the improvement brought by DA may be marginal. 

For nonlinear systems, the exact description of the transition pdf is provided by a linear partial differential equation (pde) known as the Fokker Planck Kolmogorov Equation (FPKE)~\cite{Risken1989}. Analytical solutions exist only for stationary pdf and are restricted to a limited class of dynamical systems \cite{Risken1989}. Thus researchers are looking actively at numerical approximations to solve the FPKE \cite{Kumar2007,Muscolino1997}, generally using the variational formulation of the problem. For discrete-time dynamical systems, solving for the exact solution, which is given by the Chapman-Kolmogorov Equation (CKE), yields the same problems as in the continuous-time case. Several other techniques exist in the literature to approximate the pdf evolution, the most popular being Monte Carlo (MC) methods~\cite{doucet}, Gaussian closure~\cite{GaussClosure} (or higher order moment closure), Equivalent Linearization~\cite{linear}, Stochastic Averaging~\cite{lefeb1}, Gaussian mixture approximations~\cite{Alspach1972,Ito2000,Terejanu2008}. Furthermore, all these approaches provide only an approximate description of the uncertainty propagation problem by restricting the solution to a small number of parameters.

All these assumptions employed make the problem tractable and computational efficient, which satisfies the requirement of minimizing decision latency. But the approximation given may be of little use when computing the expected loss, since the method is not sensitive to the region of interest. Such an example may be given using Monte Carlo approximations or Gaussian Sum approximations, when the propagated uncertainty offers no or very little probabilistic support in the region of interest. In other words, no particles or Gaussian components are centered in the region of interest, and even though the probabilistic support may be infinite, the expected loss computed might be underestimated.

Methods to deal with such situations have been developed, from risk sensitive filters \cite{Banavar1998} to risk sensitive particle filters \cite{Thrun2002}. The risk sensitive filters minimize the expected exponential of estimation error, controlling this way how much to weigh the higher-order moments versus the lower-order moments. While weighting more the higher-order moments, these methods are not designed to be selectively sensitive to particular regions of interest in the state space. The risk sensitive particle filter is able to generate more samples in the region of interest, but at the expense of biasing the proposal distribution, thus the particles obtained are biased towards the region of interest. While this method is appropriate for fault detection, it provides a limited output for the decision maker who is interested in querying the probability density function for different numerical quantities used in the decision process such as the expected loss, the mean or the mode of the pdf.

The present paper is concerned with providing a better approximation to the probability density function by incorporating contextual loss information held by the decision maker into the DA process. In this work we use a Gaussian mixture approximation to the probability density function. We propose a ``non-intrusive'' way in computing an approximate pdf that addresses the region of interest and it is closer to the true pdf. Non-intrusive refers here to the fact the we do not require a new DA method when incorporating the loss function into the derivation.

A progressive selection method is designed to add new Gaussian components to the initial Gaussian mixture, in assuring that probabilistic support is reaching the region of interest at the decision time. The initial weights of the new Gaussian components are set to zero and they are modified when propagated throughout the nonlinear dynamical system to minimize the error in the FPKE~\cite{Terejanu2008}. Therefore if there is any probability density mass in the region of interest it will be represented by the non-zero weight of the new Gaussian components at the decision time. 

The problem is stated in Section~\ref{sec:Problem} and the Gaussian Sum approximation to the conditional pdf is presented in Section \ref{sec:Approximation}. The progressive selection of Gaussian components is derived in Section~\ref{sec:Method}. An example to illustrate the concept is given in Section~\ref{sec:Results} and the conclusions and future work are discussed in Section~\ref{sec:Conclusion}.

%% file: problem.tex
Consider a general $n$-dimensional continuous-time noise driven nonlinear dynamic system with uncertain initial conditions and discrete measurement model, given by the equations:
\begin{align}
\dot{\textbf{x}}(t) &= \textbf{f}(t, \textbf{x}(t)) + \textbf{g}(t, \textbf{x}(t)) \bm{\Gamma}(t) \label{state_eq} \\
\textbf{z}_k &= \textbf{h}(t_k,\textbf{x}_k) + \textbf{v}_k \label{meas_eq}
\end{align}
and a set of $k$ observations, $\textbf{Z}_k = \{ \textbf{z}_i ~|~ i=1\ldots k\}$.

We denote, $\textbf{x}_k = \textbf{x}(t_k)$, $\bm{\Gamma}(t)$ represents a Gaussian white noise process with the correlation function $\mathbf{Q}\delta(t_{k+1}-t_{k})$, and the initial state uncertainty is captured by the pdf $p(t_0, \textbf{x}_0)$. The random vector $\textbf{v}_k$ denotes the measurement noise, which is temporally uncorrelated, zero-mean random sequence with known covariance, $\textbf{R}_k$. The process noise and the measurement noise are uncorrelated with each other and with the initial condition.

We are interested in finding the conditional probability density function $p(t, \textbf{x}(t) ~|~ \textbf{Z}_k)$. For $t>t_k$ we get the forecast pdf by integrating only Eq.\ref{state_eq} forward, if $t=t_k$ we are interested in the nowcast or filtered pdf and for $t<t_k$ we obtain the hindcast or the smoothed pdf.

Given a state space region of interest at a particular decision time, $t_d$, which may be represented as a \textsl{loss function} by the decision maker, $L(\textbf{x}_d,a_d)$, the expected loss of an action $a_d$ is given by:
\begin{equation}
L(a_d) = \int{L(\textbf{x}_d,a_d)p(t_d,\textbf{x}_d | \textbf{Z}_k)\mathrm{d}\textbf{x}_d}
\end{equation}

Here we will consider only the cases where $t_d > t_k$. Given approximate computational methods for the conditional pdf, $\hat{p}(t_d,\textbf{x}_d | \textbf{Z}_k)$, we are able to obtain an estimate of the expected loss and also find the optimal Bayesian decision, if a set of decisions exists.
\begin{eqnarray}
\hat{L}(a_d) &=& \int{L(\textbf{x}_d,a_d)\hat{p}(t_d,\textbf{x}_d | \textbf{Z}_k)\mathrm{d}\textbf{x}_d} \\
\hat{a}_d &=& \arg\min_{a_d} \int{L(\textbf{x}_d,a_d)\hat{p}(t_d,\textbf{x}_d | \textbf{Z}_k)\mathrm{d}\textbf{x}_d}
\end{eqnarray}

\begin{figure}
	\centering
		\includegraphics[width=3.2in]{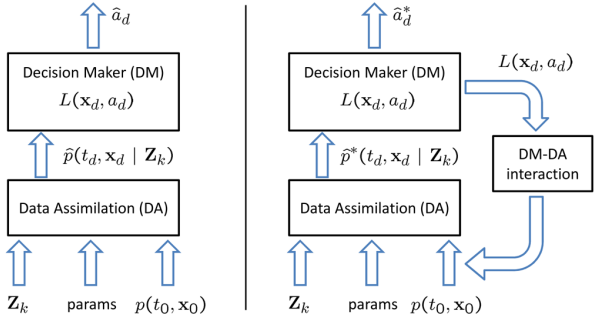}
	\caption{Left figure represents the classic approach to decision making in the data assimilation context. The right figure shows the proposed model.}
	\label{fig:figure1}
\end{figure}

The decision making process in the data assimilation framework is presented in Fig.\ref{fig:figure1}(left). Obviously if we have a good approximation for the conditional pdf in the region of interest the same can be said for the expected loss. This situation becomes more dramatic when a large deviation exists between the actual and the estimated conditional pdf in the region of interest. In the case of evaluation of a single decision, the algorithm can underestimate the actual expected loss, $\hat{L}(a_d) \ll L(a_d)$, misguiding the decision maker with respect to the magnitude of the situation. In the case when a optimal decision has to be chosen, the large difference between conditional pdfs may result in picking not only a suboptimal decision but a very consequential one.

While one can derive a new method to approximate the conditional pdf by including the loss function in the derivation and reduce the difference in the region of interest to better approximate the expected loss, it will accomplish this at the expense of worsening the approximation of the conditional pdf in the rest of the state space.

This will affect other estimates based on the conditional pdf, but the expected loss, that may be required in guiding the decision process, like the mean of the pdf, the modes of the pdf, etc. These will be biased towards the region of interest, thus misleading the decision maker.

In other words, if we call the computation of the expected loss of a given action as \textit{impact assessment} and the computation of the moments and other quantities based on the conditional pdf as \textit{situation assessment}, one will require that both to be as accurate as possible. At the limit, if we can compute exactly the conditional pdf we obtain both \textit{impact assessment} and \textit{situation assessment} since we can quantify exactly the probability of all the outcomes.

Since the decision maker holds important information regarding the use of the conditional pdf obtained from the data assimilation method, we can incorporate this information in the data assimilation process in a non-intrusive manner (do not have to derive a new method), by supplementing the inputs into the data assimilation module. The proposed method is shown in Fig.\ref{fig:figure1}(right), where a new interaction level is introduced between the decision maker and the data assimilation, that uses the contextual information provided by the decision maker to supplement the inputs of DA / change the environment in which DA is running.

Therefore we want to find an approximation to the conditional pdf, $\hat{p}^*(t_d,\textbf{x}_d | \textbf{Z}_k)$, that addresses the interest held by the decision maker and provides both a better impact and situation assessment than $\hat{p}(t_d,\textbf{x}_d | \textbf{Z}_k)$. These objectives can be captured by the following two relations:
\begin{align}
&\int \big| p(t_d,\textbf{x}_d | \textbf{Z}_k) - \hat{p}^*(t_d,\textbf{x}_d | \textbf{Z}_k) \big|^2 \mathrm{d}\textbf{x}_d \le \nonumber \\
&\quad\quad\int \big| p(t_d,\textbf{x}_d | \textbf{Z}_k) - \hat{p}(t_d,\textbf{x}_d | \textbf{Z}_k) \big|^2 \mathrm{d}\textbf{x}_d \\
&\left| \hat{L}^*(a_d) - L(a_d) \right| \le \left| \hat{L}(a_d) - L(a_d) \right|
\end{align}

In the present paper, we will design an interaction level between the decision maker and the data assimilation module that approximates the conditional pdf using a Gaussian mixture. The interaction level is adding new Gaussian components to the initial uncertainty, such that they will be positioned near the region of interest at the decision time. Their initial weights will be set to zero, thus the initial uncertainty is not changed, but the evolution of the weights is dictated by the error in the Fokker-Planck equation. Thus if any probability density mass is moving naturally towards the region of interest, the weights of the new Gaussian components will become greater than zero. Therefore the method will just make sure that if there is any probability density mass in the region of interest it will be found by the data assimilation method.

In this paper we will consider only the forecast of the conditional pdf when no measurements are available between the current time and the decision time. A suggestion, on how this can be used in the case when we have observations to assimilate between the current time and the decision time, is given in Section~\ref{sec:Method}.

%% file: approximation.tex
The nonlinear filtering problem has been extensively studied and various methods are provided in literature. The Extended Kalman Filter (EKF) is historically the first, and still the most widely adopted approach to solve the nonlinear state estimation problem. It is based on the assumption that the nonlinear system dynamics can be accurately modeled by a first-order Taylor series expansion~\cite{Crassidis2004}. Since the EKF provides us only with a rough approximation to the a posteriori pdf and solving for the exact solution of the conditional pdf is very expensive, researchers have been looking for mathematically convenient approximations.

In Ref.~\cite{Alspach1972}, a weighted sum of Gaussian density functions has been proposed to approximate the conditional pdf. The probability density function of the initial condition is given by the following Gaussian sum,
\small\begin{align}
p(t_0, \textbf{x}_0) &= \sum_{i=1}^{N} w_0^i \mathcal{N}(\textbf{x}_0 ~|~ \bm{\mu}_0^i, \textbf{P}_0^i) \label{gauss_sum} \\
\quad \mathcal{N}(\textbf{x} | \bm{\mu}, \textbf{P}) &= \left|2\pi\textbf{P}\right|^{-1/2} \exp\left[-\frac{1}{2}\left(\textbf{x}-\bm{\mu}\right)^T\textbf{P}^{-1}\left(\textbf{x}-\bm{\mu}\right)\right] \nonumber
\end{align}\normalsize
Let us assume that the underlying conditional pdf can be approximated by a finite sum of Gaussian pdfs
\begin{align}\label{gaussappr}
\hat{p}(t, \textbf{x}(t) ~|~ \textbf{Z}_k) &= \sum\limits_{i=1}^{N} w_{t|k}^i \underbrace{\mathcal{N} (\textbf{x}(t) ~|~ \bm{\mu}_{t|k}^i, \textbf{P}_{t|k}^i)}_{p_{g_i}}
\end{align}
where $\bm{\mu}_{t|k}^i$ and $\textbf{P}_{t|k}^i$ represent the conditional mean and covariance of the $i^{th}$ component of the Gaussian pdf with respect to the $k$ measurements, and $w_{t|k}^i$ denotes the amplitude of $i^{th}$ Gaussian in the mixture. The positivity and normalization constraint on the mixture pdf, $\hat{p}(t, \textbf{x}|\textbf{Z}_k)$, leads to following constraints on the amplitude vector: 
\begin{equation}\label{weightconstr}
\sum\limits_{i=1}^{N} w_{t|k}^i=1,\quad w_{t|k}^i \ge 0, \quad \forall~t
\end{equation}

A Gaussian Sum Filter \cite{Alspach1972} may be used to propagate and update the conditional pdf. Since all the components of the mixture pdf \eqref{gaussappr} are Gaussian and thus, only estimates of their mean and covariance need to be propagated between $t_k$ and $t_{k+1}$ using the conventional Extended Kalman Filter time update equations:
\small\begin{align}\label{meanprop}
\dot{\bm{\mu}}_{t|k}^i &= \textbf{f}(t,\bm{\mu}_{t|k}^i)\\\label{covprop}
\dot{\textbf{P}}_{t|k}^i &= \textbf{A}_{t|k}^i \textbf{P}_{t|k}^i+\textbf{P}_{t|k}^i (\textbf{A}_{t|k}^i)^T+\textbf{g}(t, \bm{\mu}_{t|k}^i) \textbf{Q} \textbf{g}^T(t, \bm{\mu}_{t|k}^i) \\
\textbf{A}_{t|k}^i &= \frac{\partial \textbf{f}(t, \textbf{x}(t))} {\partial \textbf{x}(t)} \bigg|_{\textbf{x}(t) = \bm{\mu}_{t|k}^i}
%
\end{align}\normalsize

In Ref.\cite{Terejanu2008} an update method of adapting the weights of different Gaussian components during propagation is given based on minimizing the error in the Fokker-Planck equation for continuous dynamical systems, and on minimizing the integral square difference between the true forecast pdf, given by the Chapman-Kolmogorov equations and its Gaussian sum approximation in the discrete time dynamical systems. 

The new weights are given by the solution of the following quadratic programing problem:
\begin{align}\label{weight_opt}
\textbf{w}_{t|k} =& \arg\min\limits_{\textbf{w}_{t|k}} ~~ \frac{1}{2} \textbf{w}_{t|k}^T(\textbf{L}+\textbf{I})\textbf{w}_{t|k} - \textbf{w}_{t|k}^T\textbf{w}_{k|k} \\
&\textrm{subject to} \quad \textbf{1}_{N\times1}^T \textbf{w}_{t|k} = 1 \nonumber \\
&\quad\quad\quad\quad\quad \textbf{w}_{t|k} \ge \textbf{0}_{N\times1} \nonumber
\end{align}
where $\textbf{w}_{t|k}\in\mathbb{R}^{N\times1}$ is a vector of Gaussian weights, and the elements of $\textbf{L}\in\mathbb{R}^{N\times N}$ are given by:
\begin{align} \label{Lmatrix}
L_{ij} &= \int\limits_V\mathbf{\mathfrak{L}}_i\mathbf{\mathfrak{L}}_{j}\textrm{d}\textbf{x} \\
\mathbf{\mathfrak{L}}_i(t, \textbf{x})&= \left[\frac{\partial p_{g_i}}{\partial\bm{\mu}^i_{t|k}}^T \textbf{f}(t,\bm{\mu}^i_{t|k}) + \sum\limits_{j=1}^{n}\sum\limits_{k=1}^{n}\frac{\partial p_{g_i}}{\partial P^{i,jk}_{t|k}}\dot{P}^{i,jk}_{t|k}\right. \nonumber\\
&\left. + \sum\limits_{j=1}^n\left(f_j(t, \textbf{x})\frac{\partial p_{g_i}}{\partial x_j} + p_{g_i}\frac{\partial f_j(t, \textbf{x})}{\partial x_j}\right.\right. \nonumber\\
&\left.\left. + \frac{1}{2}\frac{\partial d^{(1)}_j(t,\textbf{x})p_{g_i}}{\partial x_j} - \frac{1}{2}\sum\limits_{k=1}^n\frac{\partial^2d^{(2)}_{jk}(t,\textbf{x})p_{g_i}}{x_jx_k} \right)\right] \nonumber \\
%
%
d^{(1)}(t,\textbf{x}) &= \frac{1}{2}\frac{\partial \textbf{g}(t,\textbf{x})}{\partial\textbf{x}}\textbf{Q}\textbf{g}(t,\textbf{x}) \nonumber\\
d^{(2)}(t,\textbf{x}) &= \frac{1}{2}\textbf{g}(t,\textbf{x})\textbf{Q}\textbf{g}^T(t,\textbf{x}) \nonumber
\end{align}

Notice, to carry out this minimization, we need to evaluate integrals involving Gaussian pdfs over volume $V$ which can be computed exactly for polynomial nonlinearity and in general can be approximated by the Gaussian quadrature method.

The measurement update is done using Bayes rule, where the state and the covariance matrix are updated using the Extended Kalman Filter measurement update equations, and the weights are updated as it is shown in Ref.~\cite{Anderson1979}. The equations can be found in Ref.~\cite{Terejanu2008a}.








By updating the forecast weights, not only can we obtain a more accurate estimate but also a better approximation to the conditional probability density function~\cite{Terejanu2008a}. This weight update method during uncertainty propagation is very useful when the measurement model offer limited or no information in updating the states of the system.

%% file: loss.tex
The estimated conditional pdf is used to compute the expected loss. We require that the loss function provided is positive, finite everywhere and it is able to distinguish the important states from the unimportant ones. For simplicity the loss function used in this work has the following form:
\begin{equation}
L(\mathbf{x}_d,a_d) = \mathcal{N}(\textbf{x}_d ~|~ \bm{\mu}_L, \mathbf{\Sigma}_L)
\end{equation}

Observe that even with a better approximation of the weights of different Gaussian components, these components may drift away from the loss function due to first order Taylor series uncertainty propagation and limited information in the measurement update, situation which may be avoided if the conditional pdf can be found in an exact way. 

Due to the approximations used in propagating the conditional pdf it may happen that no or very little probability density mass exists in the region of interest at the decision time, depicted here by the loss function. Thus the expected loss will be significantly underestimated, misguiding this way the decision maker.

%% file: method.tex
The iterative method proposed here, is adding a set of Gaussian components to the initial pdf that are sensitive to the loss function at the decision time. After propagation, these Gaussian components will be located near the center of support of the loss function at the decision time. Initially the weights of these components are set to zero, and they will be updated in the propagation step if any probability density mass is moving in their direction. The weights at the decision time will give their relative contributions in computing the expected loss with respect to the entire conditional pdf.

An algorithm that bears similarity to the simulated annealing and the progressive correction used in particle filters~\cite{Musso2001}, is proposed in selecting the initial Gaussian components.

The main idea is to select a set of Gaussian components initially, propagate each one of them using the time update equations in the Extended Kalman Filter until the decision time is reached and based on their contributions to the expected loss, select their means and variances in the initial distribution such that after propagation the expected loss is maximized.

Let the initial pdf be given by $p(t_0, \textbf{x}_0)$ as a Gaussian sum, Eq.\ref{gauss_sum}. Compute the mean and the variance of this mixture.
\begin{align}
\bm{\mu}_0 &= \sum^{N}_{i=1} w_0^i \bm{\mu}_0^i \label{estimate} \\
\textbf{P}_0 &= \sum^{N}_{i=1} w_0^i \left[ \textbf{P}_0^i + (\bm{\mu}_0^i - \bm{\mu}_0)(\bm{\mu}_0^i - \bm{\mu}_0)^T \right]
\end{align}

Assume that we want to add another $M$ new Gaussian components to the initial pdf with zero weights and sensitive to the loss function. We sample the means of these Gaussian components from the initial distribution such that their equally weighted sum gives the mean in Eq.\ref{estimate}.
\begin{align} \label{sample_means}
\bm{\mu}_i &\sim p(t_0, \textbf{x}_0) \quad \mathrm{for} ~ i = 1 \dots M-1 \\
\bm{\mu}_M &= M\bm{\mu}_0 - \sum^{M-1}_{i=1} \bm{\mu}_i
\end{align}

The default covariance of the Gaussian components is $\textbf{D}$. We want to find the new covariance $\textbf{D}^*$ such that the covariance of the new Gaussian components matches the covariance of the initial pdf. Let $\textbf{D}^* = \gamma \textbf{D}$. Thus we want to find $\gamma$ such that we minimize the following expression:
\small\begin{eqnarray}
J_{\gamma} = \mathrm{Tr}\bigg[ \textbf{P}_0 - \frac{1}{M} \sum^{M}_{i=1} \bigg( \gamma \textbf{D} + (\bm{\mu}_i - \bm{\mu}_0)(\bm{\mu}_i - \bm{\mu}_0)^T \bigg)  \bigg] \\
\gamma = \frac{1}{\mathrm{Tr}\big[ \textbf{D} \big]} \mathrm{Tr}\bigg[ \textbf{P}_0 - \frac{1}{M} \sum^{M}_{i=1} (\bm{\mu}_i - \bm{\mu}_0)(\bm{\mu}_i - \bm{\mu}_0)^T   \bigg]
\end{eqnarray}\normalsize

Only solutions $\gamma > 0$ are accepted. Otherwise we repeat the sampling of the means, Eq. \ref{sample_means}. Once we have the initial Gaussian sum components we propagate them using the time update equations in the Extended Kalman Filter until we reach the decision time. Let $\bm{\mu}_{t_d}^i$ and $\textbf{P}_{t_d}^i$ be their means and covariances. The Gaussian components will then be weighted based on their contribution to the expected loss. A larger contribution means a more sensitive component to the loss function, thus a larger weight.

To be able to compute the weights of the Gaussian components, make sure that all of them are fairly weighted and we are not running into numerical problems and also create an indicator to mark the end of the algorithm, we compute an inflation coefficient for the loss function. Let $\mathbf{\Sigma}_L^* = \alpha \mathbf{\Sigma}_L$ be the inflated covariance of the loss function.

The inflation coefficient $\alpha$ is found such that the expected loss computed using the most distant Gaussian component from the loss function is maximized. Let the mean and the covariance of the most distant component be denoted by $\bm{\mu}_{t_d}^{max}$ and $\textbf{P}_{t_d}^{max}$ respectively. 
\begin{eqnarray}
J_{max} &=& \int \mathcal{N}(\textbf{x}_d | \bm{\mu}_L, \alpha \mathbf{\Sigma}_L) \mathcal{N}(\textbf{x}_d | \bm{\mu}_{t_d}^{max}, \textbf{P}_{t_d}^{max}) \mathrm{d}\textbf{x}_d \nonumber \\
&=& \mathcal{N}(\bm{\mu}_L ~|~ \bm{\mu}_{t_d}^{max} , \alpha \mathbf{\Sigma}_L + \textbf{P}_{t_d}^{max}) 
\end{eqnarray}

An equivalent way to seek $\alpha$ is by minimizing the negative log of the above expectation.
\small\begin{align}
&J_{min} = \mathrm{log}[\mathrm{det}(\alpha \mathbf{\Sigma}_L + \textbf{P}_{t_d}^{max})] ~+ \nonumber \\
&\quad\quad(\bm{\mu}_L - \bm{\mu}_{t_d}^{max})^T \bigg(\alpha \mathbf{\Sigma}_L + \textbf{P}_{t_d}^{max} \bigg)^{-1}(\bm{\mu}_L - \bm{\mu}_{t_d}^{max})
\end{align}\normalsize

Let us denote $\textbf{K} = \alpha \mathbf{\Sigma}_L + \textbf{P}_{t_d}^{max}$ and $\textbf{U} = (\bm{\mu}_L - \bm{\mu}_{t_d}^{max})(\bm{\mu}_L - \bm{\mu}_{t_d}^{max})^T$. We seek $\alpha > 0$ such that
\begin{align}
\frac{\partial J_{min}}{\partial \alpha} = 0 \\
\mathrm{Tr}\bigg[ \textbf{K}^{-1}\mathbf{\Sigma}_L - \textbf{K}^{-1}\textbf{U}\textbf{K}^{-1}\mathbf{\Sigma}_L \bigg] = 0 \label{my_trace}
\end{align}

After a few mathematical manipulations, Eq.\ref{my_trace} can be written in the following format:
\small\begin{align}
\mathrm{Tr}\bigg[ \textbf{K}^{-1}\mathbf{\Sigma}_L (\alpha\textbf{I} + \textbf{P}_{t_d}^{max}\mathbf{\Sigma}_L^{-1} - \textbf{U}\mathbf{\Sigma}_L^{-1}) \textbf{K}^{-1}\mathbf{\Sigma}_L \bigg] = 0 \label{new_trace}
\end{align}\normalsize

Denote $\textbf{A} = \textbf{K}^{-1}\mathbf{\Sigma}_L$ and $\textbf{B} = \alpha\textbf{I} + \textbf{P}_{t_d}^{max}\mathbf{\Sigma}_L^{-1} - \textbf{U}\mathbf{\Sigma}_L^{-1}$. Observe that for $\alpha > 0$ the matrix $\textbf{A}$ is symmetric and positive definite. 

\textbf{\textit{Lemma}}: ~ If $\mathrm{Tr}[\textbf{A}\textbf{B}\textbf{A}]=0$ and $\textbf{A}$ is symmetric and positive definite then $\mathrm{Tr}[\textbf{B}]=0$.

\textit{Proof}: ~ Let $\textbf{A} = \textbf{V}\textbf{S}\textbf{V}^T$ be a singular value decomposition of matrix $\textbf{A}$, where $\textbf{V}$ is a unitary matrix and $\textbf{S}$ is a diagonal matrix. Our trace can now be written as $\mathrm{Tr}[\textbf{A}\textbf{B}\textbf{A}] = \mathrm{Tr}[\textbf{V}\textbf{S}\textbf{V}^T\textbf{B}\textbf{V}\textbf{S}\textbf{V}^T] = \mathrm{Tr}[\textbf{S}^2\textbf{B}]$.

If $\mathrm{Tr}[\textbf{S}^2\textbf{B}]=0$ then $\textbf{S}^2\textbf{B}$ is a commutator. Thus there is $\textbf{X}$ and $\textbf{Y}$ such that $\textbf{S}^2\textbf{B} = \textbf{X}\textbf{Y} - \textbf{Y}\textbf{X}$. But $\textbf{B} = \textbf{S}^{-2}\textbf{X}\textbf{Y} - \textbf{S}^{-2}\textbf{Y}\textbf{X} = \textbf{X}^*\textbf{Y} - \textbf{Y}\textbf{X}^*$, where $\textbf{X}^* = \textbf{S}^{-2}\textbf{X}$. Therefore $\textbf{B}$ is also a commutator, hence $\mathrm{Tr}[\textbf{B}]=0$.

Applying the previous lemma to Eq.\ref{new_trace} we get
\begin{align}
\mathrm{Tr}\bigg[ \alpha\textbf{I} + \textbf{P}_{t_d}^{max}\mathbf{\Sigma}_L^{-1} - \textbf{U}\mathbf{\Sigma}_L^{-1} \bigg] = 0 \label{new_new_trace}
\end{align}

Therefore we accept solutions $\alpha > 1$ that satisfy the following relation
\begin{align}
\alpha = \frac{1}{n}\mathrm{Tr}\bigg[ (\textbf{U} -\textbf{P}_{t_d}^{max})\mathbf{\Sigma}_L^{-1} \bigg]
\end{align}

For $\alpha \le 1$ we stop the algorithm. Otherwise we continue getting the weights of the Gaussian components by solving the following optimization problem:
\begin{align}
\textbf{w} =& \arg\min\limits_{\textbf{w}} ~~ \frac{1}{2} \textbf{w}^T \textbf{M} \textbf{w} - \textbf{w}^T \textbf{N} \\
&\textrm{subject to} \quad \textbf{1}_{M\times1}^T \textbf{w} = 1 \\
& \hspace{.7in} \textbf{w} \ge \textbf{0}_{M\times1}
\end{align}
where $\textbf{w} \in \mathbb{R}^{M\times 1}$, $\textbf{M} \in \mathbb{R}^{M\times M}$ and $\textbf{N} \in \mathbb{R}^{M\times 1}$ and their entries are given by:
\begin{eqnarray}
m_{ij} = \mathcal{N}\bigg\{\bm{\mu}_{t_d|0}^j ~\bigg|~ \bm{\mu}_{t_d|0}^i ~,~ \textbf{P}_{t_d|0}^i+\textbf{P}_{t_d|0}^j\bigg\} \\
n_{i} = \mathcal{N}\bigg\{\bm{\mu}_L ~\bigg|~ \bm{\mu}_{t_d|0}^i ~,~ \textbf{P}_{t_d|0}^i+\mathbf{\Sigma}_L^*\bigg\}
\end{eqnarray}

The new pdf used to sample the new means is given by:
\begin{align}
p_{new}(t_0, \textbf{x}_0) = \sum^M_{i=1} \mathcal{N}(\textbf{x}_0 ~|~ \bm{\mu}_i, \beta\textbf{D}^*)
\end{align}

\begin{algorithm}[H]
\algsetup{indent=1.2em}
\caption{Progressive Selection of Gaussian Components}
\label{prog_sel}
\begin{algorithmic}[1]
\REQUIRE $t_d$ - decision time \\
$p(t_0, \textbf{x}_0)$ - initial probability density function \\ 
$M$ - number of extra Gaussian components \\
$\textbf{D}$ - Gaussian component covariance \\
$w_\mathrm{tol}$ - add only Gaussian components with weights greater than this threshold \\
$L(\textbf{x})$ - loss function $\mathcal{N}\{\textbf{x}|\bm{\mu}_L,\mathbf{\Sigma}_L\}$ \\
\medskip
\STATE $\hat{p}_0 = p(t_0, \textbf{x}_0)$, $\alpha = \infty$, $\gamma = -1$
\WHILE{$(\alpha > 1) ~\&~ \mathrm{maxiter}$}
	\STATE The mean and the covariance of the initial pdf \\
		$\bm{\mu}_0 = \sum^{N}_{i=1} w_0^i \bm{\mu}_0^i$ \\
		$\textbf{P}_0 = \sum^{N}_{i=1} w_0^i \left[ \textbf{P}_0^i + (\bm{\mu}_0^i - \bm{\mu}_0)(\bm{\mu}_0^i - \bm{\mu}_0)^T \right]$
	\WHILE{$(\gamma < 0)$}
		\STATE Get the means of the Gaussian components \\
			Draw $\bm{\mu}_i \sim p(t_0, \textbf{x}_0) \quad \mathrm{for} ~ i = 1 \dots M-1$ \\
			Set $\bm{\mu}_M = M\bm{\mu}_0 - \sum^{M-1}_{i=1} \bm{\mu}_i$
		\STATE \small$\gamma = \frac{1}{\mathrm{Tr}\big[ \textbf{D} \big]} \mathrm{Tr}\bigg[ \textbf{P}_0 - \frac{1}{M} \sum^{M}_{i=1} (\bm{\mu}_i - \bm{\mu}_0)(\bm{\mu}_i - \bm{\mu}_0)^T \bigg]$\normalsize
	\ENDWHILE
	\STATE Get the covariance of the Gaussian components \\	
	$\textbf{P}_0^i = \gamma\textbf{D}$
	\STATE Propagate the moments from $t=0$ to $t=t_d$\\
		$\dot{\bm{\mu}}_{t|0}^i = \textbf{f}(t,\bm{\mu}_{t|0}^i)$\\
		$\dot{\textbf{P}}_{t|0}^i = \textbf{A}_{t|0}^i \textbf{P}_{t|0}^i+\textbf{P}_{t|0}^i (\textbf{A}_{t|0}^i)^T+\textbf{g} \textbf{Q} \textbf{g}^T$	
	\STATE Get the most distant component \\
	  by computing the Mahanalobis distance \\
		\small$d_i = (\bm{\mu}_L - \bm{\mu}_{t_d|0}^i)^T \bigg(\textbf{P}_{t_d|0}^i+\mathbf{\Sigma}_L\bigg)^{-1}(\bm{\mu}_L - \bm{\mu}_{t_d|0}^i)$\normalsize \\
		$\bm{\mu}_{t_d|0}^{max} ~,~ \textbf{P}_{t_d|0}^{max} = \arg\max(d_i)$
	\STATE Compute optimal $\alpha$ and the inflated matrix $\Sigma_L^*$ \\
		\footnotesize$\alpha = \frac{1}{n}\mathrm{Tr}\bigg[ \bigg((\bm{\mu}_{t_d|0}^{max}-\bm{\mu}_L)(\bm{\mu}_{t_d|0}^{max}-\bm{\mu}_L)^T - \textbf{P}_{t_d|0}^{max} \bigg) \Sigma_{L}^{-1} \bigg]$\normalsize
	\STATE \textbf{if} $\alpha < 1$ \textbf{then} $\alpha = 1$ \textbf{end if} \\
	$\Sigma_L^* = \alpha \Sigma_L$
	\STATE Elements of $\textbf{M} \in \mathbb{R}^{M\times M}$ and $\textbf{N} \in \mathbb{R}^{M\times 1}$ \\
	$m_{ij} = \mathcal{N}\bigg\{\bm{\mu}_{t_d|0}^j ~\bigg|~ \bm{\mu}_{t_d|0}^i ~,~ \textbf{P}_{t_d|0}^i+\textbf{P}_{t_d|0}^j\bigg\}$ \\
	$n_{i} = \mathcal{N}\bigg\{\bm{\mu}_L ~\bigg|~ \bm{\mu}_{t_d|0}^i ~,~ \textbf{P}_{t_d|0}^i+\mathbf{\Sigma}_L^*\bigg\}$ \\	
	\STATE Compute the weights \\
	$\textbf{w} = \arg\min\limits_{\textbf{w}} ~~ \frac{1}{2} \textbf{w}^T \textbf{M} \textbf{w} - \textbf{w}^T \textbf{N}$ \\
	\hspace{.3in}$\textrm{subject to} \quad \textbf{1}_{M\times1}^T \textbf{w} = 1$ \\
	\hspace{1in}$\textbf{w} \ge \textbf{0}_{M\times1}$
	\STATE Set $\hat{p}_0 = \sum^M_{j=1} w_j \mathcal{N}\{\textbf{x}|\bm{\mu}^j_0,\beta\mathbf{P}^j_0\}$
\ENDWHILE
\STATE Set $p_{\textrm{NEW}}(t_0, \textbf{x}_0) = p(t_0, \textbf{x}_0) ~ + $ \\
\quad\quad $\sum^{M, w_j \ge w_\mathrm{tol}}_{j=1} 0 \times  \mathcal{N}\{\textbf{x}|\bm{\mu}^j_0,\mathbf{P}^j_0\}$
\RETURN $p_{\textrm{NEW}}(t_0, \textbf{x}_0)$
\end{algorithmic}
\end{algorithm}

Where $\beta \le 1$ is a coefficient that controls the decrease of the initial variance. If $\alpha$ has decreased from the previous iteration this means that the Gaussian components are getting closer to the loss function and therefore we can decrease the variance of the initial distribution to finely tune the position of the Gaussian components, otherwise $\beta = 1$. We continue to sample new means from the new pdf until $\alpha < 1$ or the maximum number of time steps has been reached. The entire algorithm is presented in Table~\ref{prog_sel}.

%% file: meas.tex
While not the scope of this paper, the above method can also be applied when measurements are available between the current time and the decision time. The progressive selection algorithm will be applied every time a measurement has been assimilated and the a posteriori pdf has been found. The drawback of this procedure is that the number of Gaussian components will increase linearly with the number of measurements. Better ways to deal with the measurement updates are set as future work.

In the case of multiple loss functions, the algorithm is run once for each one of the loss functions, creating sets of initial Gaussian components sensitive to their loss function.

%% file: results.tex

\begin{figure*}[!t]
\centering
\subfigure[TRUTH: Numerical approximation FPKE]{\includegraphics[width=2.2in]{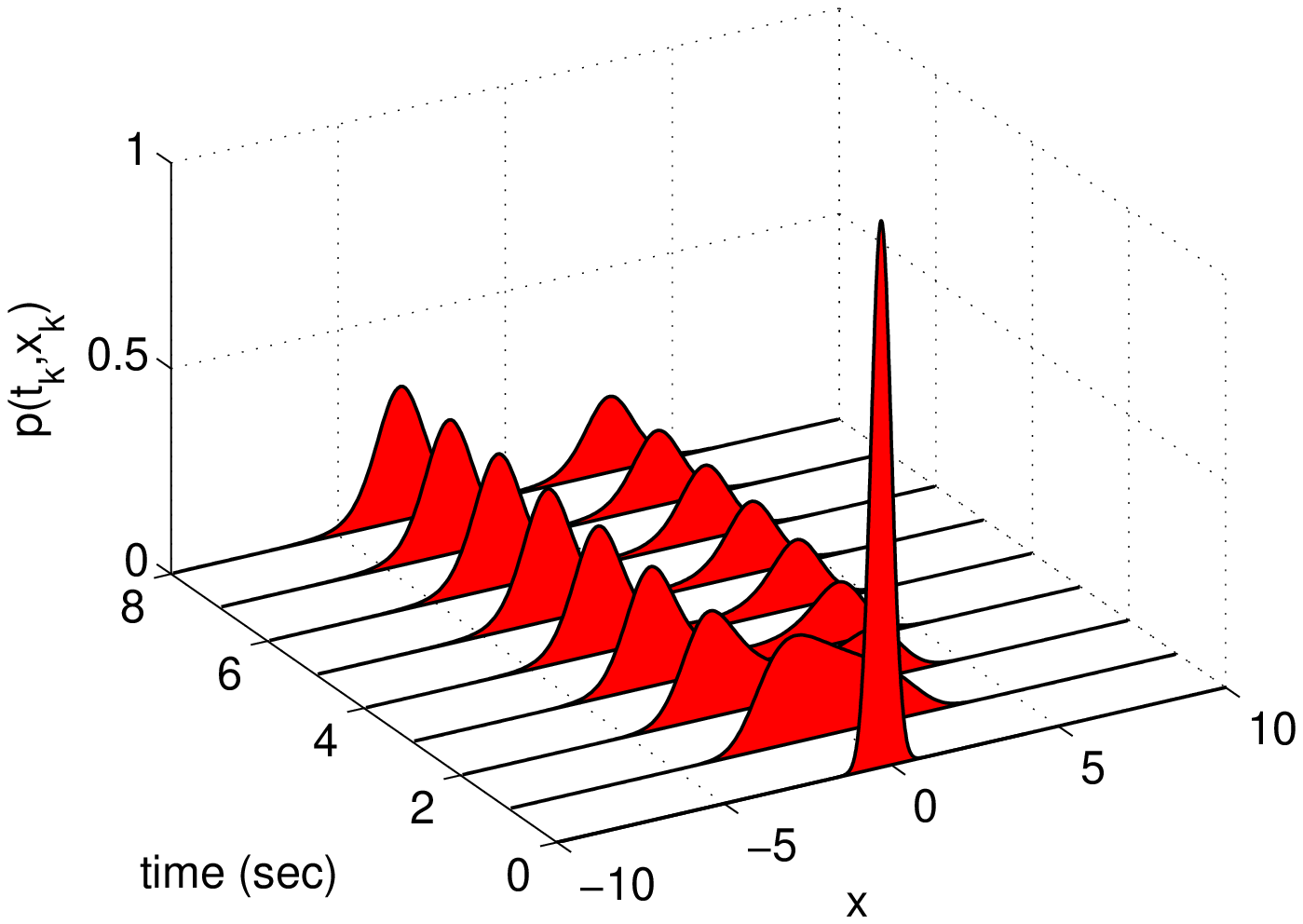}} 
\subfigure[EKF: first order Taylor expansion approximation]{\includegraphics[width=2.2in]{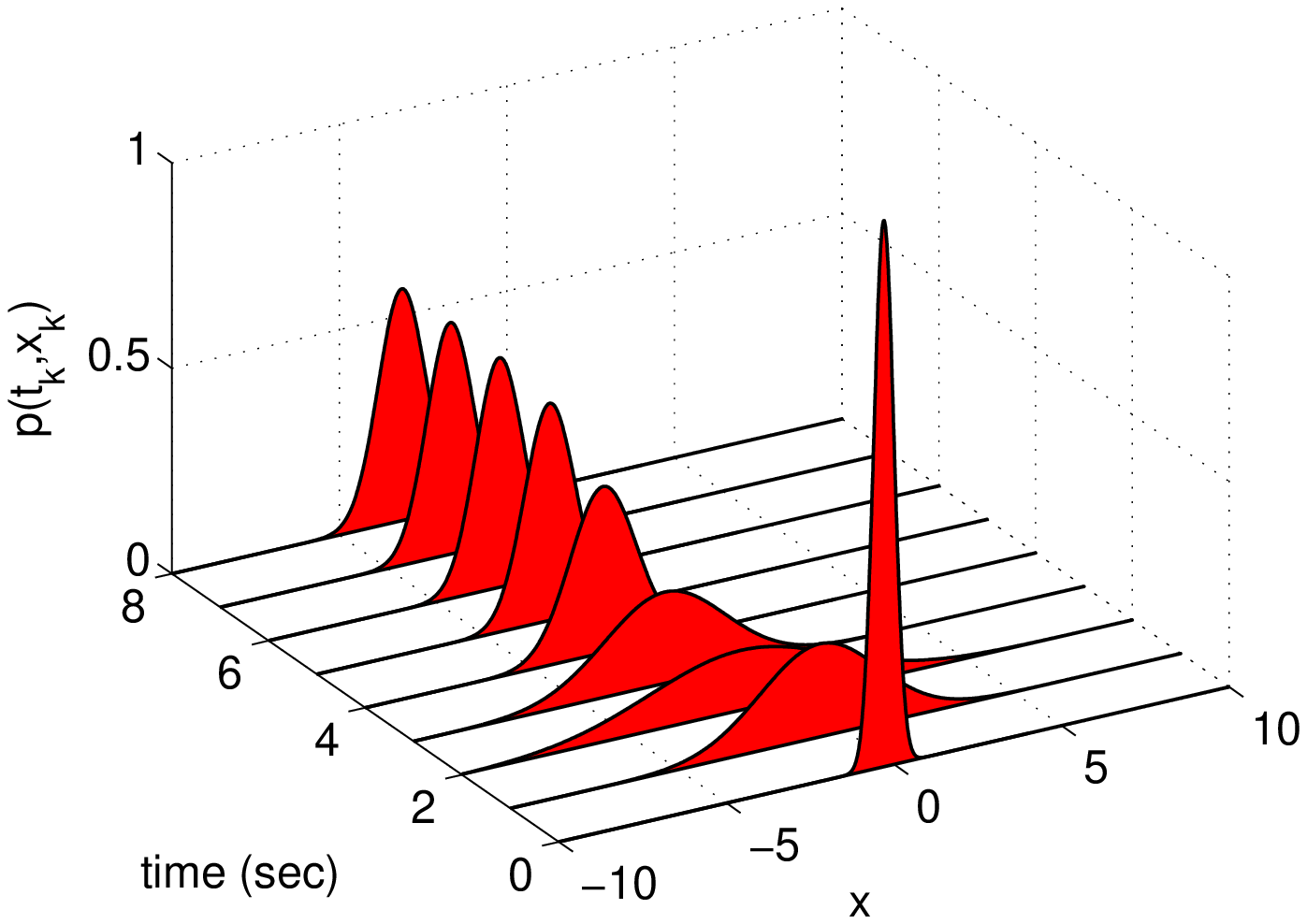}} 
\subfigure[GS\_BCK: back propagated means]{\includegraphics[width=2.2in]{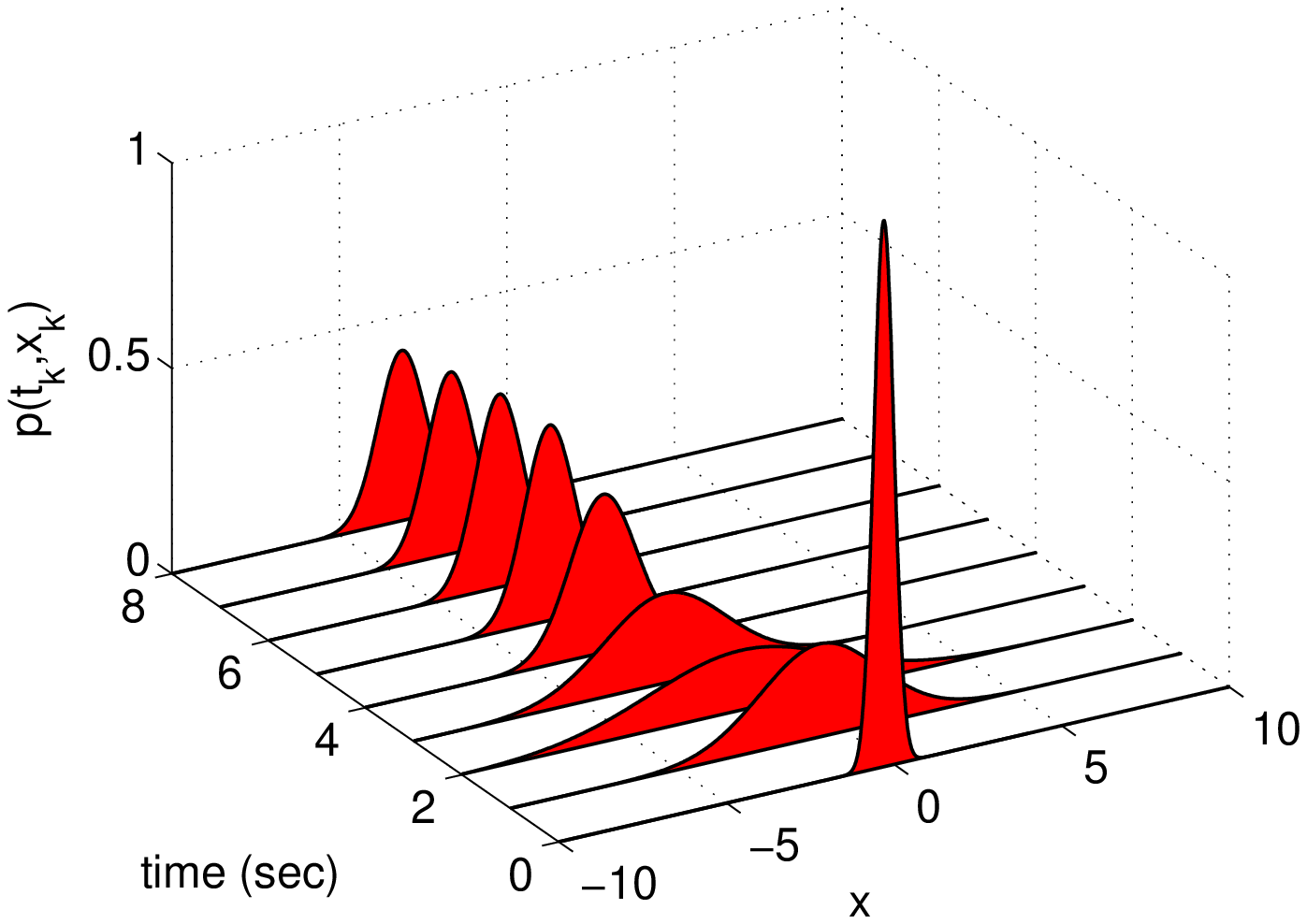}}
\subfigure[GS\_DEC: progressive selection of Gaussian components]{\includegraphics[width=2.2in]{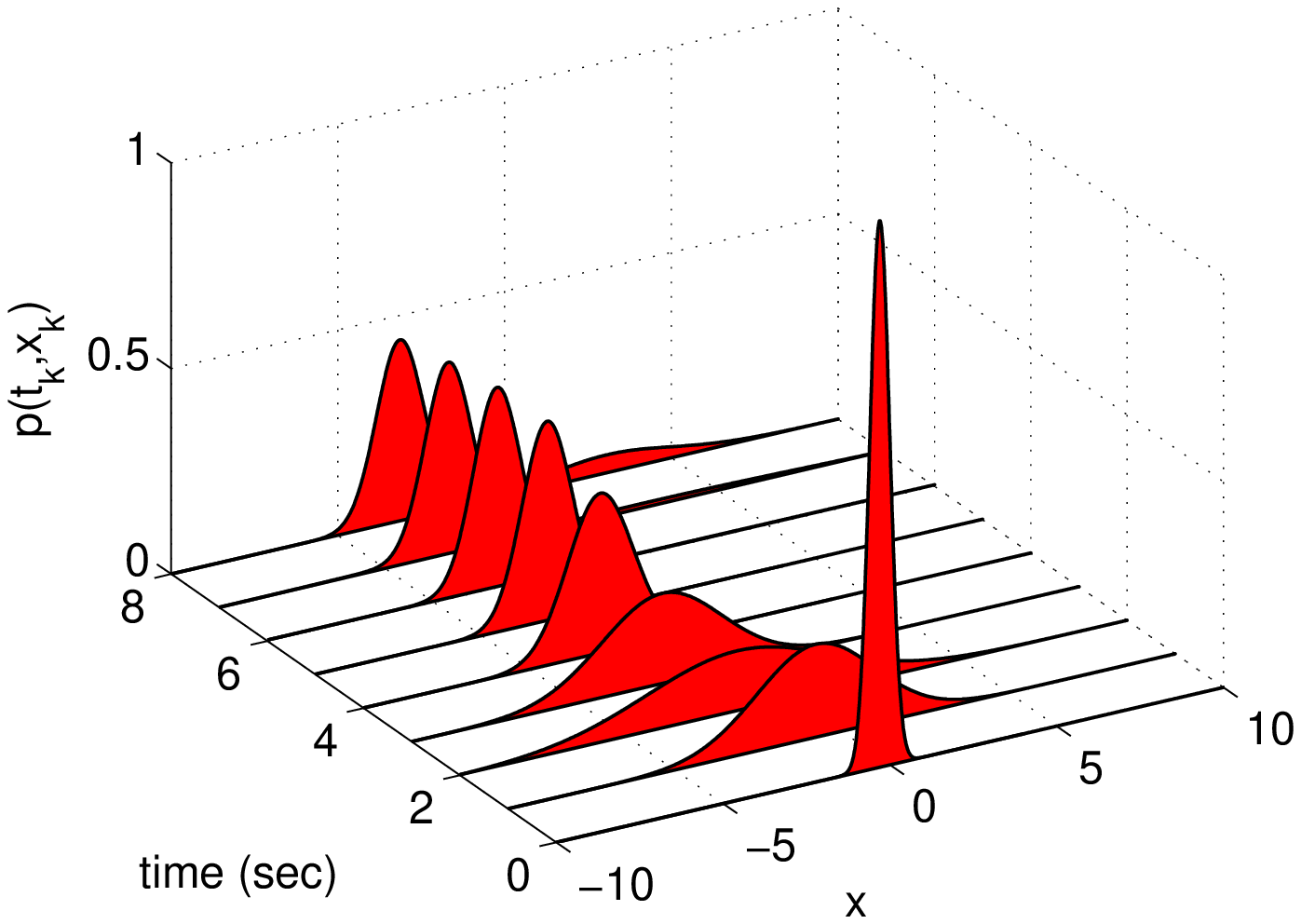}} 
\subfigure[Probability density function at $t_k = 8$~sec]{\includegraphics[width=2.2in]{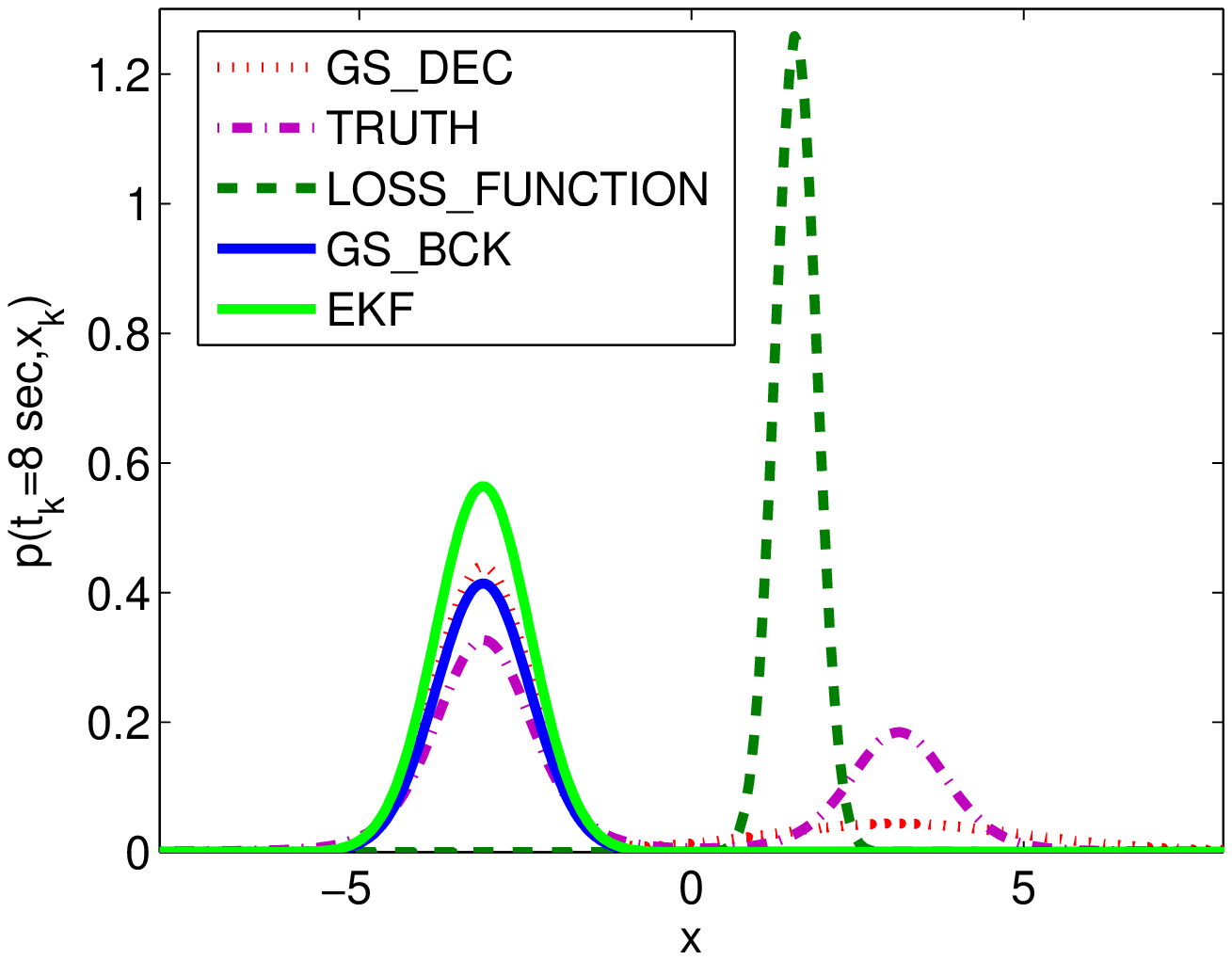}}
\subfigure[The evolution of the pdf used to sample the means of the Gaussian components]{\includegraphics[width=2.2in]{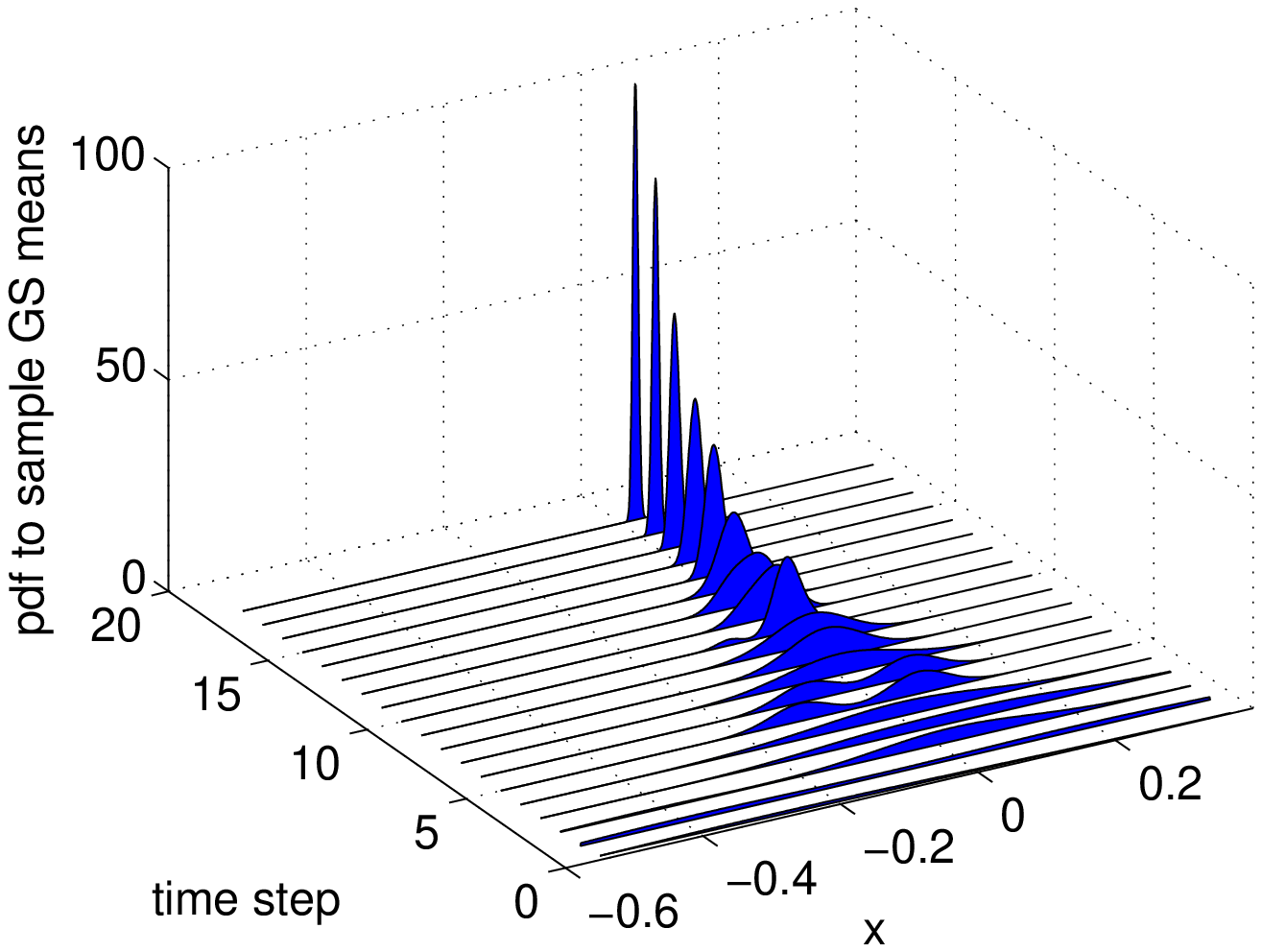}}
\caption{The evolution of the Forecast pdf and the Sampling pdf}\label{fig:results}
\end{figure*}

To illustrate the concept of incorporating contextual information into the uncertainty propagation algorithm, we consider the following continuous-time dynamical system with uncertain initial condition given by \eqref{continuous_ic}:
\begin{align}\label{continuous_ic}
\dot{x} &= \textrm{sin}(x) +  \Gamma(t) \quad \textrm{where} \quad Q = 1\\
x_0 &\sim \mathcal{N}(-0.3, 0.3^2) \nonumber
\end{align}

The state space region of interest is depicted by the following loss function, and the time of decision is at $t_d = 8$~sec.
\begin{equation}
L(x) = \mathcal{N}(x ~|~ \frac{\pi}{2} ~,~ 0.1^2)
\end{equation}

First we compute an accurate numerical solution based on the FPKE, and this will stand as the true probability density function. The performance measures for this method will be labeled as TRUTH. The evolution of the pdf using this method can be seen in Fig.\ref{fig:results}a. 

Three other approximations for the pdf are provided including the method presented in this paper. The first approximation propagates the initial uncertainty using the first order Taylor expansion, Eq.\eqref{meanprop} and Eq.\eqref{covprop}, also known as the Extended Kalman Filter time update equations, labeled later as EKF. The evolution of the pdf for this method is presented in Fig.\ref{fig:results}b. 

For the next approximation method, we add another $5$ Gaussian components to the initial one, creating this way a Gaussian mixture with $6$ components. The means of the new components are just the result of back propagation (from $t_d = 8$~sec to $t_c = 0$~sec) of $5$ equidistant samples taken in the $3$ sigma bound of the loss function support. The variance of the new components is set to $10^{-10}$ and their initial weights are set to zero. The label used for this method is GS\_BCK and the evolution of the pdf is shown in Fig.\ref{fig:results}c. While all the means of the new Gaussian components are positioned in the loss function support region, their variances are very being difficult to see the probability density mass in that region.

We apply the method presented in this paper to generate at most $5$ new Gaussian components to be added to the initial condition. Their means and variances are returned by the progressive selection algorithm, Alg.$1$. The initial weights of the new Gaussian components have also been set to zero. The default value for the $\beta$ coefficient is $0.9$ and Gaussian components are included only if their weights are greater than $w_{tol} = 10^{-3}$. The label used for this method is GS\_DEC and its corresponding pdf is presented in Fig.\ref{fig:results}d.

The evolution the Gaussian components for the last two methods is also achieved using the first-order Taylor expansion, but it is interrupted every $\Delta t = 0.5$~sec to adjust the weights of different Gaussian components using the optimization in Eq.\eqref{weight_opt}.  

The following performance measures have been computed for the methods used in the experiment:
\small
\begin{eqnarray}
\hat{L}_d &=& \int L(x) \hat{p}(t_d,x_d) \mathrm{d}x_d \\
\hat{R}_{err} &=& \frac{1}{L_d} \left| L_d - \hat{L}_d \right| \\
ISD &=& \int \big| p(t_d,x_d) - \hat{p}(t_d,x_d) \big|^2 \mathrm{d}x_d \\
WISD &=& \int L(x) \big| p(t_d,x_d) - \hat{p}(t_d,x_d) \big|^2 \mathrm{d}x_d
\end{eqnarray}
\normalsize

In Fig.\ref{fig:results}e it is plotted the forecast pdf at time $t_d$ for all the methods. Our method, GS\_DEC, is able to better estimate the probability density mass in the region of interest.
%

In Table I, we present the performance measures after $500$ Monte Carlo runs. The expected loss given by the GS\_DEC method is consistently better approximated over all the Monte Carlo runs than the EKF and the GS\_BCK method. We also are able to consistently give a better approximation to the pdf overall and also in the region of interest than the EKF method, which justify the use of this method. Compared with the GS\_BCK we do a better job in average in approximating the pdf which suggests that there is a trade off in selecting the Gaussian components regarding their means and variances.

In Fig.~\ref{fig:results}f it is plotted the evolution of the pdf used to sample the means of the new Gaussian components for one particular run. The pdf used in the first iteration is our initial uncertainty and we see how it converges, as the number of iterations increases, to a particular region in the state space that is sensitive to the loss function at the decision time.
%

%

\begin{table}[htp]
  \centering
  \caption{Performance measures - 500 Monte Carlo runs}
    \begin{tabular}{rrrrr}
    \toprule
          & $\hat{L}_d$    & $\hat{R}_{err}$ & $ISD$   & $WISD$ \\
    \midrule
    TRUTH & 0.0332 & N/A   & N/A   & N/A \\
    EKF   & 4.93E-09 & 1.0000 & 0.1840 & 0.0015 \\
    GS\_BCK & 0.0001 & 0.9968 & 0.0536 & 0.0015 \\
    GS\_DEC (mean) & 0.0256 & 0.2300 & 0.0470 & 0.0004 \\
    \midrule
    \multicolumn{ 5}{r}{GS\_DEC: Percentile Table - 500 Observations} \\
    Percent & $\hat{L}_d$    & $\hat{R}_{err}$ & $ISD$   & $WISD$ \\
    \midrule
    0.0\% & 0.0010 & 0.0151 & 0.0368 & 0.0002 \\
    5.0\% & 0.0142 & 0.0230 & 0.0378 & 0.0003 \\
    10.0\% & 0.0177 & 0.0271 & 0.0380 & 0.0003 \\
    25.0\% & 0.0229 & 0.0566 & 0.0387 & 0.0003 \\
    50.0\% & 0.0257 & 0.2270 & 0.0491 & 0.0003 \\
    75.0\% & 0.0313 & 0.3090 & 0.0514 & 0.0004 \\
    90.0\% & 0.0323 & 0.4670 & 0.0574 & 0.0006 \\
    95.0\% & 0.0324 & 0.5710 & 0.0601 & 0.0007 \\
    100.0\% & 0.0327 & 0.9700 & 0.0705 & 0.0014 \\
    \bottomrule
    \end{tabular}
  \label{tab:addlabel}
\end{table}

%% file: conclusion.tex
An interaction level between the decision maker and the data assimilation module has been designed, such that we can incorporate contextual information held by the decision maker into the data assimilation process to better approximate the conditional probability function.

The progressive selection algorithm is run once at the beginning of the simulation to supplement the initial uncertainty with new Gaussian components that are sensitive to the loss function at the decision time. The weights of all the Gaussian components are then updated during the propagation based on the Fokker Planck Equation. This way we obtain not only a better approximation of the probability density function in the region of interest but also a better approximation overall.

The cost of this overall improvement is an increase in the number of Gaussian components. The principal benefit is not the modest increase in accuracy overall, but the significantly enhanced accuracy within the decision maker's region of interest.